\documentclass[12pt]{article}
\usepackage{mymacros}

\title{Classical and quantum curves of 5d Seiberg's theories \\and their 4d limit}

\makeatletter
\renewcommand\@date{{
  \vspace{-\baselineskip}
  \large\centering
  \begin{tabular}{@{}c@{}}
    Oleg Chalykh\textsuperscript{$a$} 
  \end{tabular}
  \!,
  \begin{tabular}{@{}c@{}}
    Yongchao L\"u\textsuperscript{$b$} 
  \end{tabular}
  \bigskip

  {\small \textsuperscript{$a$}\,School of Mathematics, 
	University of Leeds, 
	Leeds, LS2 9JT, UK } \\
    \normalsize {o.chalykh@leeds.ac.uk} \par
 {\small
	 \textsuperscript{$b$}\,School of Physics, Korea Institute for Advanced Study, Seoul 02455, Korea} \\
    \normalsize {yongchaolu@kias.re.kr}
}}
\makeatother

\begin{document}
\maketitle
\begin{abstract}
In this work, we examine the classical and quantum Seiberg--Witten curves of 5d $\mathcal{N}=1$ SCFTs and their 4d limits. The 5d theories we consider are Seiberg's theories of type $E_{6,7,8}$, which serve as the UV completions of 5d SU(2) gauge theories with 5, 6, or 7 flavors. Their classical curves can be constructed using the five-brane web construction \cite{KimYagi14}. We also use it to re-derive their quantum curves \cite{Moriyama2020_SpectralTSDelPezzo}, by employing a $q$-analogue of the Frobenius method in the style of \cite{MoriyamaY2021_AffineWeylQuantumCurve}. This allows us to compare the reduction of these 5d curves with the 4d curves, i.e. Seiberg--Witten curves of the Minahan--Nemeschansky theories and their quantization, which have been identified in \cite{ACLccrgRankone} with the spectral curves of rank-1 complex crystallographic elliptic Calogero-Moser systems.
\end{abstract}
\tableofcontents

\section{Introduction}

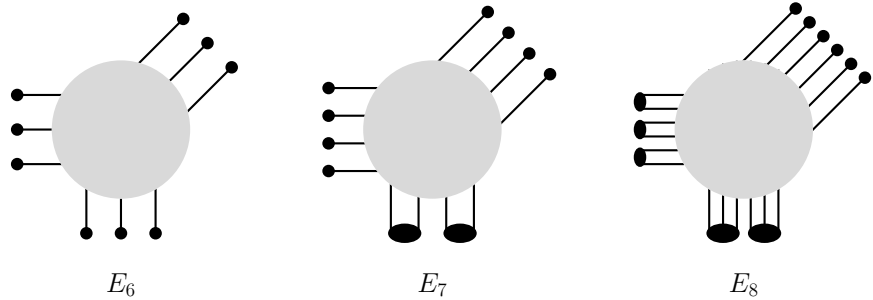
\begin{figure}[htbp]
\centering   
\begin{tikzpicture}[thick,scale=0.46, every node/.style={scale=0.5}]
\begin{scope}[xshift=0cm]
\draw (1.0,-3) -- (1.0,0);
\draw (0.0,-3) -- (0,0);
\draw (-1.0,-3) -- ( -1.0,0);
\draw[fill] (1.0,-3) circle (0.15);
\draw[fill] (0,-3) circle (0.15);
\draw[fill] (-1.0,-3) circle (0.15);

\draw (-3,-1.0) -- (0, -1.0);
\draw (-3,0.0) -- (0, 0);
\draw (-3,1.0) -- (0, 1.0);
\draw[fill] (-3,-1.0) circle (0.15);
\draw[fill] (-3,0) circle (0.15);
\draw[fill] (-3,1.0) circle (0.15);

\draw (1.8,3.2) -- (-0.7,0.7);
\draw (2.5,2.5) -- (0,0);
\draw (3.2,1.8) -- (0.7,-0.7);
\draw[fill] (1.8,3.2) circle (0.15);
\draw[fill] (2.5,2.5) circle (0.15);
\draw[fill] (3.2,1.8) circle (0.15);

\fill[fill = white] (0,0) circle (2);
\fill[gray!30] (0,0) circle (2);
    \node[anchor=north] at (0,-4){\LARGE $E_6$};
\end{scope}

\begin{scope}[xshift=9cm]
    \draw (1.2,-3) -- (1.2, 0);
\draw (0.4,-3) -- (0.4, 0);
\draw (-0.4,-3) -- (-0.4, 0);
\draw (-1.2,-3) -- ( -1.2, 0);
\draw[fill] (0.8,-3) ellipse (0.45cm and 0.25 cm);
\draw[fill] (-0.8,-3) ellipse (0.45cm and 0.25 cm);

\draw (-3,-1.2) -- (0, -1.2);
\draw (-3,-0.4) -- (0, -0.4);
\draw (-3,0.4) -- (0, 0.4);
\draw (-3,1.2) -- (0, 1.2);
\draw[fill] (-3,-1.2) circle (0.15);
\draw[fill] (-3,-0.4) circle (0.15);
\draw[fill] (-3,0.4) circle (0.15);
\draw[fill] (-3,1.2) circle (0.15);

\draw (3.4,1.6) -- (0.9,-0.9);
\draw (2.8,2.2) -- (0.3,-0.3);
\draw (2.2,2.8) -- (-0.3,0.3);
\draw (1.6,3.4) -- (-0.9,0.9);
\draw[fill] (3.4,1.6) circle (0.15);
\draw[fill] (2.8,2.2) circle (0.15);
\draw[fill] (2.2,2.8) circle (0.15);
\draw[fill] (1.6,3.4) circle (0.15);

\fill[fill = white] (0,0) circle (2);
\fill[gray!30] (0,0) circle (2);
    \node[anchor=north] at (0,-4){\LARGE $E_7$};
\end{scope}

\begin{scope}[xshift= 18cm]
    \draw (-3,-1.0) -- (-1.732,-1.0);
    \draw (-3,-0.6) -- (-1.908,-0.6);
    \draw (-3,-0.2) -- (-1.99,-0.2);
    \draw (-3,0.2) -- (-1.99,0.2);
    \draw (-3,0.6) -- (-1.908,0.6);
    \draw (-3,1.0) -- (-1.732,1.0);
    \draw[fill] (-3,-0.8) ellipse (0.15cm and 0.25cm);
    \draw[fill] (-3,0.0) ellipse (0.15cm and 0.25cm);
    \draw[fill] (-3,0.8) ellipse (0.15cm and 0.25cm);

    \draw (1.0,-3) -- (1.0,1.732);
    \draw (0.6,-3) -- (0.6,1.908);
    \draw (0.2,-3) -- (0.2,1.99);
    \draw (-0.2,-3) -- (-0.2,1.99);
    \draw (-0.6,-3) -- (-0.6,1.908);
    \draw (-1.0,-3) -- (-1.0,1.732);
    \draw[fill] (-0.6,-3) ellipse (0.45cm and 0.25cm);
    \draw[fill] (0.6,-3) ellipse (0.45cm and 0.25cm);

    \draw (3.5,1.5) -- (1.0,-1.0);
    \draw (3.1,1.9) -- (0.6,-0.6);
    \draw (2.7,2.3) -- (0.2,-0.2);
    \draw (2.3,2.7) -- (-0.2,0.2);
    \draw (1.9,3.1) -- (-0.6,0.6);
    \draw (1.5,3.5) -- (-1.0,1.0);

    \draw[fill] (3.5,1.5) circle (0.15);
    \draw[fill] (3.1,1.9) circle (0.15);
    \draw[fill] (2.7,2.3) circle (0.15);
    \draw[fill] (2.3,2.7) circle (0.15);
    \draw[fill] (1.9,3.1) circle (0.15);
    \draw[fill] (1.5,3.5) circle (0.15);

    \fill[fill = white] (0,0) circle (2);
    \fill[gray!30] (0,0) circle (2);
    \node[anchor=north] at (0,-4){\LARGE $E_8$};
\end{scope}




\end{tikzpicture}
\caption{Shown above are the five-brane webs corresponding to 5d Seiberg's theories. The internal part of each diagram is schematically represented by a central circle, with the external legs illustrated in detail. The black dots represent seven-branes, and the lines represent five-branes. }
\label{fig:5dweb}
\end{figure}

In this paper, we review and compare the classical and quantum Seiberg--Witten (SW) curves of certain 5d and 4d SCFTs. 
The 5d theories we consider are the rank-1 Seiberg theories of type $E_{6,7,8}$ \cite{Seiberg1996_5dSUSY}, and their 4d limit corresponds to the rank-1 Minahan--Nemeschansky (MN) theories \cite{Minahan:1996fg, Minahan:1996cj}. The classical SW curves for the MN theories have been recently related \cite{ACLccrgRankone} to complex crystallographic elliptic Calogero--Moser systems and elliptic pencils of a rather special form, admitting a natural quantization. On the other hand, the classical curves for 5d rank-1 Seiberg's theories on $\mathbb{R}^4 \times S^1$ are also given by particular rational elliptic fibrations (see \cite{ClossetM_4dKKtheories} for a recent comprehensive review). Their description in terms of canonical Weierstrass models has been established in \cite{MinahanNWEn, EguchiSakaiEString1, EguchiSakaiEString2}\footnote{It is worth mentioning that the 5d and 4d theories we consider can be obtained as suitable limits of the 6d E-string theory \cite{GanorMS1996_braneCYT26dE8, MinahanNWEn, EguchiSakaiEString1}.}.  However, for us it is much more convenient to use their derivation based on five-brane webs \cite{AharonyHK_web5dSUSY,KimYagi14}. The schematic representation of the webs for Seiberg's theories is shown in Figure \ref{fig:5dweb}. 

Note that the five-brane web presentation is not unique, and different presentations are connected through Hanany-Witten transitions \cite{HananyWitten96}.  As explained in \cite{KimYagi14}, the curves derived from different diagrams are related via suitable birational coordinate transformations, realizing Hanany--Witten transitions at the level of 5d SW curves. 
The web diagrams shown in Figure \ref{fig:5dweb}, first introduced in \cite{BBT09}, are particularly well-suited for passing to a 4d limit.

As the 5d SW curves are embedded in the complex torus \(\mathbb{C}^{\ast} \times \mathbb{C}^{\ast}\), their quantization can be realized within the quantum torus in terms of $q$-difference operators \cite{Moriyama2020_SpectralTSDelPezzo, MoriyamaY2021_AffineWeylQuantumCurve}. As our main results, we performed a systematic derivation of the 5d quantum curves using the web diagram in Figure \ref{fig:5dweb}, and we proposed two types of 4d limits that exactly reproduce the quantum curves of the 4d theories presented in \cite{ACLccrgRankone}.

\section{4d classical curves} 
Here we review the properties of the SW curves for the rank-1 Minahan--Nemeschansky theories of type $E_{6,7,8}$, following \cite{ACLccrgRankone} where further details can be found. The SW integrable systems of these theories have been identified with certain complex crystallographic elliptic Calogero--Moser systems. The general construction of these integrable systems is due to \cite{EFMV11ecm}. We need a special case, associated with  elliptic curves with $\Z_m$ symmetry where $m=3,4$ and $6$ for the cases $E_6, E_7$ and $E_8$, respectively. Namely, we consider $\E=\C/\Z\omega_1+\Z\omega_2$ with $\omega_2/\omega_1=e^{\mathrm{i}\pi/3}$ for $m=3$ or $6$, and $\omega_2/\omega_1=e^{\mathrm{i}\pi/2}$ for $m=4$. The relevant integrable systems with one degree of freedom are described by a hamiltonian $h=h(p,q)$ on $\mathrm{T}^*\E$, of the form
\begin{equation}\label{h}
    h(p,q)=p^m+A_2(q)p^{m-2} +\dots + A_m(q)\,,
\end{equation}
with $A_i(q)$ elliptic functions of $q\in \E$. In fact, we have multi-parametric families of such hamiltonians, with $6,7$ and $8$ parameters for the types $E_6, E_7$ and $E_8$, respectively. 
Their explicit expressions are a bit complicated to reproduce here. They become more transparent when written in invariant coordinates, see below.  

According to \cite{ACLccrgRankone}, the hamiltonian dynamics governed by $h(p,q)$ admits a Lax representation with a Lax matrix $L=L(p,q; \alpha)$ of size $m$, depending on the spectral parameter $\alpha\in\E$. The classical spectral curves are found from the characteristic polynomial of $L$,
\begin{equation*}
    \det(L(\alpha)-k\Id)=0\,,\qquad \alpha\in\E\,,\ k\in\C\,.
\end{equation*}
These are $m$-sheeted branched coverings of $\E$ of the form
\begin{equation*}
    k^m+B_2k^{m-2} +\dots + B_m=0\,,\qquad B_i=B_i(p,q;\alpha).
\end{equation*}
These curves are invariant under the hamiltonian flow, hence the coefficients $B_i$ are functions of $\alpha$ and $z=h(p,q)$ only. There is a peculiar duality between the level sets of the hamiltonian $h$ and the spectral curves.
Namely, as shown in \cite{ACLccrgRankone}, the spectral curves can be written as
\begin{equation}\label{specd}
    \Ssigma=\{(k,\alpha)\,:\,h^\vee(k, \alpha)=z \}\,,
\end{equation}
parameterised by the value $z$ of the hamiltonian $h(p,q)$. Here $h^\vee$ is obtained from \eqref{h} by replacing $p\mapsto k$, $q\mapsto \alpha$ and switching to the ``dual'' parameters. The precise formula for these dual parameters can be found in \cite{ACLccrgRankone}; it will not be important for us here. Due to the $\mathbb{Z}_m$-symmetry of $h$, the spectral curves $\Ssigma$ are also invariant under the $\Z_m$-action
\begin{equation*}\label{aced}
    k\to\omega k\,,\quad \alpha\to\omega^{-1}\alpha\,,\qquad \omega=e^{2\pi i/m}\,,
\end{equation*}
so we may consider their quotient $\Sigma:=\Ssigma/\Z_m$. Both $\Ssigma$ and $\Sigma$ compactify to smooth curves, and while $\Ssigma$ has a fairly high genus, $g(\Ssigma)=m^2+1$, the quotient curve is elliptic, i.e. $g(\Sigma)=1$. As a result, the spectral curves $\Sigma$, as $z$ varies, define an elliptic fibration on $\mathrm{T}^*\E/\Z_m$ which is identified with the SW fibration, with $z$ viewed as the Coulomb branch (CB) modulus. The SW differential is induced by the Liouville 1-form $\lambda=k\mathrm{d}\alpha$. These fibrations admit a nice geometric interpretation as certain elliptic pencils in $\mathbb P^2$.
To see this, it is convenient to rewrite the spectral curves in $\Z_m$-invariant coordinates
\begin{equation}
    x=u(\alpha)\,,\qquad y=v(\alpha)k\,,
\end{equation}
where $u, v$ are the functions from Table \ref{table1}.
\begin{table}[h]
\centering
\begin{align*}
\renewcommand\arraystretch{1.6}
\begin{array}{c | c c c }
\hline
     m   & 3 & 4 & 6 \\
    \hline
    \omega_2/\omega_1 & e^{{ \pi i }/{3}} & i &  e^{{ \pi i }/{3}} \\
    \hline
    u & \frac12\wp'(\alpha) & \wp^2(\alpha) & \wp^3(\alpha) \\
    \hline
    v &  \wp(\alpha) & \frac12\wp'(\alpha) & \frac12\wp(\alpha) \wp'(\alpha) \\
    \hline
\end{array}
\end{align*}
 \caption{The elliptic functions $u,v$.}
\label{table1}
\end{table}

The canonical Poisson bracket $\{k, \alpha\}=1$ induces the bracket $\{y,x\}=m(x-e_1)(x-e_2)$, with suitable $e_1, e_2\in\C$. By shifting and rescaling $x$, and rescaling the bracket, we can always bring it into the form
\begin{equation}\label{poisson}
\{y,x\}=x(x-1)\,.    
\end{equation}
The Seiberg--Witten differential in $x,y$ coordinates takes the form
\begin{equation}\label{sw22}
 \lambda=\frac{y\,\mathrm{d}x}{x(x-1)}\,.
\end{equation}  
Finally, the fibration \eqref{specd} in $x,y$ coordinates takes the form
\begin{equation}\label{pen}
    Q(x,y)-zP(x)=0\,,
\end{equation}
where $P$ and $Q$ are given below, case by case. 

\paragraph{Case $m=3$, type $E_6$\,:}
Here $P=x(x-1)$ and 
$Q=y^3+Q_2y+Q_3$, 
\begin{equation}\label{cc3}
    Q=y^3+\left(a_2 x(x-1)-b_2(x-1)+c_2x \right)y+a_3 x \left(x-1\right)^2-b_3\left(x-1\right) + c_3x\,.
\end{equation}

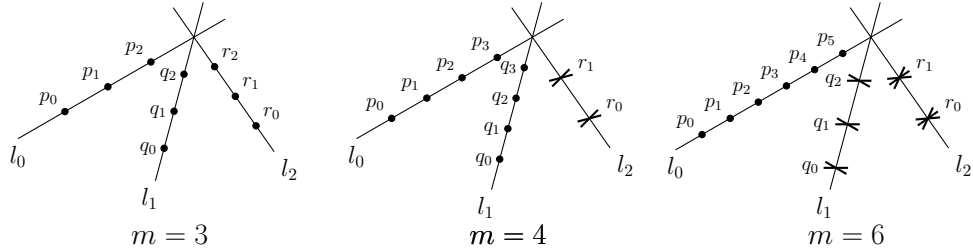
\begin{figure}[H]
\centering
\begin{tikzpicture}[scale=0.6, every node/.style={scale=0.65}]

\begin{scope}[xshift=22.5 cm]
\draw (30:0.8) -- (30+180:5.);
\draw (75:0.8) -- (75+180:3.5);
\draw (125:0.8) -- (125+180:3);
\node[anchor=north] at (30+180:5.){\large $l_0$};
\node[anchor=north east, xshift = 0.2cm] at (75+180:3.5){\large $l_1$};
\node[anchor=north west] at (125+180:3.){\large $l_2$};

\foreach \i in {1,...,6}
{\pgfmathtruncatemacro{\y}{(6 - \i )};
\filldraw[black] (30+180: 0.72* \i ) circle (2pt) node[sloped, anchor=south east]{$p_{\y}$};}
\foreach \i in {1,...,3}
{\pgfmathtruncatemacro{\y}{(3 - \i )};
\draw (75+180: 1.0*\i ) node[sloped, xshift = -0.2cm,anchor= east]{$q_{\y}$};
\draw (75+180: 1.0*\i ) pic [thick, rotate=-15] {cross2};
}
\foreach \i in {1,...,2}
{\pgfmathtruncatemacro{\y}{(2 - \i )};
\draw (125+180: 1.1*\i ) node[sloped, xshift = 0.2cm,anchor= south west]{$r_{\y}$};
\draw (125+180: 1.1*\i ) pic [thick, rotate=35] {cross3};
}
\node[anchor=north,xshift = -0.5 cm] at (0,-4){\Large $m=6$};
\end{scope}

\begin{scope}[xshift=15 cm]
\draw (30:0.8) -- (30+180:4.5);
\draw (75:0.8) -- (75+180:3.5);
\draw (125:0.8) -- (125+180:3.);
\node[anchor=north] at (30+180:4.5){\large $l_0$};
\node[anchor=north east, xshift = 0.2cm] at (75+180:3.5){\large $l_1$};
\node[anchor=north west] at (125+180:3.){\large $l_2$};

\foreach \i in {1,...,4}
{\pgfmathtruncatemacro{\y}{(4 - \i )};
\filldraw[black] (30+180: 0.9* \i ) circle (2pt) node[sloped, anchor=south east]{$p_{\y}$};
\filldraw[black] (75+180: 0.7* \i ) circle (2pt) node[sloped, anchor=east]{$q_{\y}$};
}
\foreach \i in {1,...,2}
{\pgfmathtruncatemacro{\y}{(2 - \i )};
\draw (125+180: 1.1*\i ) node[sloped, xshift = 0.2cm,anchor= south west]{$r_{\y}$};
\draw (125+180: 1.1*\i ) pic [thick,rotate=30] {cross2};

\node[anchor=north,xshift = -0.5 cm] at (0,-4){\Large $m=4$};
}

\end{scope}

\begin{scope}[xshift=7.5 cm]
\draw (30:0.8) -- (30+180:4.5);
\draw (75:0.8) -- (75+180:3.3);
\draw (125:0.8) -- (125+180:3.1);
\node[anchor=north] at (30+180:4.5){\large $l_0$};
\node[anchor=north east, xshift = 0.2cm] at (75+180:3.3){\large $l_1$};
\node[anchor=north west] at (125+180:3.1){\large $l_2$};

\foreach \i in {1,...,3}
{\pgfmathtruncatemacro{\y}{(3 -\i)};
\filldraw[black] (30+180: 1.1* \i ) circle (2pt) node[sloped, anchor=south east]{$p_{\y}$};
\filldraw[black] (75+180: 0.85* \i ) circle (2pt) node[sloped, anchor=east]{$q_{\y}$};
\filldraw[black] (125+180: 0.8* \i ) circle (2pt) node[sloped, anchor=south west]{$r_{\y}$};
}
\node[anchor=north,xshift = -0.5 cm] at (0,-4){\Large $m=3$};
\end{scope}

\end{tikzpicture}
\caption{Elliptic pencils. We use black dots to represent simple base points, 2-crosses for double points, and 3-crosses for triple points. In homogeneous coordinates, $x=U/W$, $y=V/W$ and the lines are $l_0: W=0$, $l_1: U=0$, $l_2: U-W=0$. The positions of the points are related to the mass parameters of the SCFT.}
\label{fig:pencil}
\end{figure}

Denoting $f=Q-zP$, it is easy to observe the following conditions:
\begin{align}
\begin{split}
    x, y \to \infty :& \quad f\propto y^3+a_2x^2y+a_3x^3  \\ 
    x\to 0 : & \quad f\propto y^3+b_2y+b_3 \\
    x \to 1 : & \quad f\propto y^3+c_2y+c_3\,. 
\end{split}
\end{align}
This means that in this case \eqref{pen} describes a pencil of cubics in $\mathbb P^2$ passing through $9$ points located on three lines $x=0, 1, \infty$ as shown in Figure \ref{fig:pencil}.  

\paragraph{Case $m=4$, type $E_7$\,:}
Here $P=x(x-1)^2$ and $Q=y^4+Q_2y^2+Q_3y +Q_4$, where 
\begin{align}
\begin{split} \label{cc4}
   Q_2 & = a_2 x\left(x-1\right)-b_2 \left(x-1\right)+2 c_2 x, \\
   Q_3 & = a_3x\left(x-1\right)^2 + b_3(x-1)^2, \\
   Q_4 & = a_4 x^2 \left(x-1\right){}^2+(a_2-b_2)c_2x(x-1)+b_4(x-1)^2+c_2^2x^2\,.
\end{split}
   \end{align}
In this case $f=Q-zP$ has the following properties:  
\begin{align}
\begin{split}
    x, y \to \infty :&  \quad f\propto y^4+a_2x^2y^2+a_3x^3y+a_4x^4 \\
    x\to 0 : & \quad f\propto y^4+b_2y^2+b_3y+b_4 \\
    x \to 1 : & \quad f\propto (y^2+c_2)^2\\
    & \quad \frac{\partial f}{\partial x} \propto y^2+c_2\,.
\end{split}
\end{align}
Hence, \eqref{pen} describes in this case a pencil of quartics (of geometric genus 1) passing through $10$ points, two of which are ordinary double points, as indicated in Figure \ref{fig:pencil}.

\paragraph{Case $m=6$, type $E_8$\,:}
Here $P=x^2(x-1)^3$ and $Q = y^6+Q_2y^4+Q_3y^3+Q_4y^2+Q_5y+Q_6$, where
\begin{align}
\begin{split} \label{cc6}
   Q_2 =&  a_2 x \left(x-1\right)-2b_2\left(x-1\right)+3 c_2x, \\
   Q_3 =& a_3 x\left(x-1\right)^2 +2 b_3\left(x-1\right)^2, \\
   Q_4 =&  a_4 x^2 \left(x-1\right)^2-a_2 b_2x\left(x-1\right)^2+2 a_2 c_2 x^2\left(x-1\right) \\
   & \quad -b_2 c_2 x\left(x-1\right)(x+3)  +b_2^2\left(x-1\right){}^2 +3 c_2^2 x^2, \\
   Q_5  =& \big(a_5 x^2 \left(x-1\right) +(a_2b_3 -a_3 b_2)x \left(x-1\right) \\ 
   & \quad + a_3 c_2 x^2  -b_3 c_2 x\left(x-3\right) 
   -2 b_2 b_3 \left(x-1\right)\big)\left(x-1\right)^2,  \\
   Q_6  =& a_6 x^2 \left(x-1\right){}^4+\left(c_2 \left(a_4-\left(a_2-b_2\right) \left(b_2-c_2\right)\right)-a_3 b_3+b_3^2\right) x^2 \left(x-1\right)^2
   \\
   & \quad +\left(c_2^2 \left(a_2-2 b_2+c_2\right)+a_3 b_3-2 b_3^2\right)x\left(x-1\right)^2 \\
   & \qquad +\left(c_2^2 \left(a_2-2 b_2+c_2\right)+b_3^2\right)x \left(x-1\right) -b_3^2\left(x-1\right) + c_2^3x^2\,. 
\end{split}
\end{align}
In this case $f=Q-zP$ has the following properties:  
\begin{align}
\begin{split}
    x, y \to \infty :&  \quad f\propto y^6+a_2x^2y^4+a_3x^3y^3+a_4x^4y^2+a_5x^5y+a_6x^5 \\
    x\to 0 : & \quad f\propto (y^3+b_2y+b_3)^2 \label{eq:6b}\\
    & \quad \frac{\partial f}{\partial x} \propto y^3+b_2y+b_3 \\
    x \to 1 : & \quad f\propto (y^2+c_2)^3 \\
    & \quad \frac{\partial f}{\partial x} \propto (y^2+c_2)^2 \\
    & \quad \frac{\partial^2 f}{\partial x^2} \propto y^2+c_2\,. 
\end{split}
\end{align}
Hence, \eqref{pen} describes in this case a pencil of sextics (of geometric genus 1) passing through $11$ points, three of which are ordinary double points, and two are ordinary triple points, see Figure \ref{fig:pencil}.

The above parameters $a_i, b_i, c_i$ are symmetric combinations of the mass parameters of the SCFT. In each case, one attaches mass parameters $\lambda_{j}$, $\mu_{j}$ and $\nu_{j}$ to the points $x=\infty, 0$ and $1$, respectively. Their geometric meaning is that they are the residues of the SW differential \eqref{sw22} on each of the curves of the pencil \eqref{pen} (independently of the value of $z$). Then 
\begin{equation}\label{symmetric masses}
a_i=\sigma_i(\lambda_j)\,,\quad b_i=\sigma_i(\mu_j)\,,\quad c_i=
\sigma_i(-\nu_j)\,,    
\end{equation}
where $\sigma_i$ denotes the $i$th elementary polynomial. The mass parameters are assumed normalised so that $a_1=b_1=c_1=0$. We refer to $a_i, b_i, c_i$ as symmetric masses.

In the $m=3$ case, we have $3$ mass parameters attached to each of the points, so we have $6$ symmetric mass parameters $a_{2,3}, b_{2,3}, c_{2,3}$. 

In the $m=4$ case, we have $4$ mass parameters attached to each of $x=\infty$ and $x=0$, and further $2$ masses attached to $x=1$, hence we have $7$ symmetric masses $a_{2,3,4}$, $b_{2,3,4}$, $c_{2}$. 

In the $m=6$ case, we have $6$ masses attached to $x=\infty$, $3$ masses at $x=0$, and further $2$ at $x=1$. This gives $8$ symmetric masses $a_{2,3,4,5, 6}$, $b_{2,3}$, $c_{2}$.

\section{5d classical curves and 4d limits}
The 5d SW curves for the brane webs shown in Figure \ref{fig:5dweb} have been studied in \cite{KimYagi14}. Here we reproduce their results and further investigate two specific types of 4d limits. We then check that they match the 4d curves from the previous section.

\subsection{5d classical curves}
Here we review the construction of the classical SW curves of the 5d theories following \cite{KimYagi14}. The 5d curve is defined by an equation $g(t,w)=0$ in $\mathbb{C}^\ast \times \mathbb{C}^\ast $ with
\begin{align}
    g(t,w) = \sum_{\substack{p \ge 0, q \ge 0 \\ p+q \le m }} c_{p,q} t^p w^q\,,\qquad m=3,4,6. 
\end{align}
Here, as before, we will associate $m=3,4,6$ with the cases $E_6, E_7$ and $E_8$, respectively. The SW differential $\lambda$ is determined (up to an exact 1-form) by the log-canonical 2-form
\begin{align} \label{clift_1}
    d \lambda = d \, \mathrm{log} \,t \wedge d\, \mathrm{log}\, w.
\end{align}
As we will see, the conditions imposed on $g(t,w)$ in \cite{KimYagi14} will be in a clear parallel with the 4d case.

\paragraph{Case $m=3$, type $E_6$:}
In that case \cite{KimYagi14} imposes the following conditions:
\begin{align} \label{5dcd3}
\begin{split} 
    w, t \to \infty : & \quad \sum_{p=0}^3 c_{p, 3-p} t^p w^{3-p} = c_{0,3} \prod_{i=0}^2 (w + L_i t) \\ 
    t\to 0 : & \quad \sum_{q=0}^3 c_{0,q} w^q = c_{0,3} \prod_{i=0}^2 (w-M_i)\\
    w \to 0 : & \quad \sum_{p=0}^3 c_{p,0} t^p = c_{3,0} \prod_{i=0}^2 (t-N_i) \,.
\end{split}
\end{align}
Here $L_i, M_i, N_i$ are the mass parameters. From \eqref{5dcd3}, we get
\begin{align}\label{eqn:3vertex}
    c_{3,0} = c_{0,3} \prod_{i=0}^2 L_i, \quad c_{0,0} = -c_{0,3} \prod_{i=0}^2 M_i, \quad  c_{0,0} = -c_{3,0} \prod_{i=0}^2 N_i
\end{align}
which imply a compatibility condition
\begin{align}
    \prod_{i=0}^2 M_i = \prod_{i=0}^2 L_i \prod_{i=0}^2 N_i\,.
\end{align}
We may further impose $\prod_{i=0}^{2} M_i =1$ and $\prod_{i=0}^2 N_i =1$ (hence, $\prod_{i=0}^2 L_i =1$ as well), by rescaling $t$ and $w$. This reduces the number of independent mass parameters to $6$, the same number as in the 4d case.  

After that, the function $g$ is uniquely determined, up to an overall factor and one free parameter, $U$, identified with the CB modulus. The result is
\begin{align} \label{5dcc3}
   g(t,w)=&  w^3 + \left(\chi_1^\lambda t - \chi_1^{\mu} \right) w^2 + \left( \chi_2^{\lambda} t^2 + U t + \chi_2^{\mu} \right) w + (t^3-\chi_1^{\nu}t^2+\chi_2^\nu t-1)\,.
\end{align}
Here and below we use the notation $\chi_j^\lambda$ ($\chi_j^\mu$ and $\chi_j^\nu$, respectively) for the $j$th elementary symmetric polynomial of $L_i$ ($M_{i}$ and $N_{i}$, respectively).

\paragraph{Case $m=4$, type $E_7$:}
In this case, the prescriptions in \cite{KimYagi14} are as follows:
\begin{align}
\begin{split}
    w, t \to \infty :&  \sum_{p=0}^4 c_{p, 4-p} t^p w^{4-p} = c_{0,4} \prod_{i=0}^3 (w + L_i t) \\ 
    t\to 0 : & \sum_{q=0}^4 c_{0,q} w^q = c_{0,4} \prod_{i=0}^3 (w-M_i)\\
    w \to 0 : &  \sum_{p=0}^4 c_{p,0} t^p = c_{4,0}  (t-N_0)^2(t-N_1)^2 \\
              & \sum_{p=0}^3 c_{p,1} t^p \propto  (t-N_0)(t-N_1)\,.
\end{split}
\end{align}
We have a compatibility condition
\begin{align}
    \prod_{i=0}^3 M_i = \prod_{i=0}^3 L_i \prod_{i=0,1} N_i^2\,.
\end{align}
We may further impose $\prod_{i=0}^3 L_i = \prod_{i=0}^{3} M_i=\prod_{i=0,1} N_i =1$, reducing the number of independent mass parameters to $7$. This determines the function $g$ uniquely, up to an overall factor and the CB modulus, $U$. The result is
\begin{align} \label{5dcc4}
\begin{split}
   g(t,w)=&  w^4 + \left(\chi_1^{\lambda } t - \chi_1^{\mu} \right) w^3 + \left( \chi_2^{\lambda} t^2 + U t + \chi_2^{\mu} \right) w^2\\
   & \quad + \left( t^2 - \chi_1^{\nu}t + 1 \right) \left( \chi_3^{\lambda} t - \chi_3^{\mu}\right) w + \left( t^2-\chi_1^{\nu } t+1 \right)^2\,.
\end{split}
\end{align}

\paragraph{Case $m=6$, type $E_8$:}
In this case, \cite{KimYagi14} prescribes the following conditions:
\begin{align} 
\begin{split}
    w, t \to \infty :&  \sum_{p=0}^6 c_{p, 6-p} t^p w^{6-p} = c_{0,6} \prod_{i=0}^5 (w + L_i t) \\ 
    t\to 0 : & \sum_{q=0}^6 c_{0,q} w^q = c_{0,6} \prod_{i=0}^2 (w-M_i)^2 \\
             & \sum_{q=0}^5 c_{1,q} w^q \propto \prod_{i=0}^2 (w-M_i) \\
    w \to 0 : &  \sum_{p=0}^6 c_{p,0} t^p = c_{6,0} (t-N_0)^3(t-N_1)^3 \\
              & \sum_{p=0}^5 c_{p,1} t^p \propto  (t-N_0)^2(t-N_1)^2 \\
              & \sum_{p=0}^4 c_{p,2} t^p \propto  (t-N_0)(t-N_1)\,.
\end{split}
\end{align}
These imply the compatibility condition
\begin{align}
    \prod_{i=0}^2 M_i^2 = \prod_{i=0}^5 L_i \prod_{i=0,1} N_i^3\,.
\end{align}
We further impose $\prod_{i=0}^5 M_i =\prod_{i=0}^2 L_i=\prod_{i=0,1} N_i =1$, by rescaling $t$ and $w$. Altogether this reduces the number of mass parameters to $8$. This determines $g$ uniquely, up to a factor and a free CB modulus, $U$. The final answer is 
\begin{align} \label{5dcc6}
\begin{split}
g(t,w) = & w^6 + \left(  \chi_1^{\lambda} t - 2 \chi_1^{\mu} \right) w^5+\left( \chi _2^{\mu } t^2 - \left(\chi_1^{\lambda} \chi_1^{\mu}+\chi _1^{\nu } \chi_2^{\mu }+\chi_5^{\lambda }\right) t+\left(\chi_1^{\mu }\right)^2+2 \chi_2^{\mu } \right) w^4\\
&  + \left( \chi_3^{\lambda } t^3 + U t^2+ \left(\chi_1^{\mu } \chi _2^{\mu } \chi_1^{\nu } + 3 \chi_1^{\nu } + \chi_1^{\lambda} \chi_2^{\mu} + \chi_1^{\mu} \chi _5^{\lambda}\right) t - 2 \chi_1^{\mu} \chi_2^{\mu} - 2\right) w^3\\
& \quad + \left(t^2 - \chi_1^{\nu } t+1\right) \left(\chi_4^{\lambda } t^2 - \left(\chi_1^{\lambda } + \chi_1^{\mu } \chi_1^{\nu } + \chi_2^{\mu } \chi_5^{\lambda } \right) t + \left(\chi_2^{\mu}\right)^2+2 \chi_1^{\mu } \right) w^2\\
& \qquad + \left(t^2-\chi_1^{\nu } t+1\right)^2 \left( \chi_5^{\lambda} t - 2 \chi_2^{\mu} \right) w+\left(t^2-\chi_1^{\nu } t+1\right)^3\,. 
\end{split}
\end{align}

\subsection{4d limit}
We would like to obtain the 4d curves as a limit of the 5d ones. To achieve this, we make the substitution \cite{BBT09}\footnote{The following redefinition 
\begin{align}\label{clift_2}
   t = x , \quad w = (1-x)\, \exp (\frac{\beta y}{1-x})
\end{align}
also works and will give the same results \cite{KimYagi14}.
}

\begin{align}\label{clift_1}
    t = x e^{\beta y}, \quad w = (1-x) e^{\beta y}
\end{align}
and we relate the 5d mass parameters $L_i, M_i, N_i$ to the 4d mass parameters, denoted $l_i, m_i, n_i$, as follows
\begin{align}
  L_i = e^{\beta \lambda_i}, \quad  M_i = e^{-\beta \mu_i}, \quad  N_i = e^{\beta \nu_i}\,,
\end{align}
with $\sum \lambda_i=\sum \mu_i=\sum \nu_i=0$. We also make a substitution for the CB modulus $U$ as
\begin{align}\label{Uexp}
    U = \sum_{k=0}^\infty u_k \beta^k\,, 
\end{align}
with $u_k$ yet to be chosen. We can find that 
Note that the log-canonical 2-form transforms under \eqref{clift_1} into
\begin{align}
   d \lambda = d \, \mathrm{log} \,t \wedge d\, \mathrm{log}\, w= \beta \frac{dx \wedge dy}{x(x-1)} ,
\end{align}
in agreement with \eqref{sw22}.

\paragraph{Case $m=3$, type $E_6$:}
We make the choice
\begin{align}
    u_0 = -6,\, u_1 = 0,\,u_2 = a_2+b_2+c_2\,, 
\end{align}
where $a_i$, $b_i$, $c_i$ are $i$-th symmetric polynomials of $\lambda_k$, $\mu_k$, $\nu_k$, respectively. Then a computer check using Mathematica shows that the $\beta^3$ term of the expansion of $g(t,w)$ gives the 4d curve \eqref{cc3}, with $u_3$ related to the CB modulus $z$ by a suitable shift. 

\paragraph{Case $m=4$, type $E_7$:}
We make the choice
\begin{align}
    u_0 = -12,\, u_1 = 0,\,u_2 = 2 \left(a_2+b_2+c_2\right),\, u_3  = 0\,, 
\end{align}
where $a_i$, $b_i$, $c_i$ are $i$-th symmetric polynomials of $\lambda_k$, $\mu_k$, $\nu_k$, respectively. The 4d curve then appears as the $\beta^4$ term of the expansion of $g(t,w)$, and it matches with the curve \eqref{cc4} upon when $u_4$ is identified with $z$ upon suitable shift.

\paragraph{Case $m=6$, type $E_8$:}
With the choice
\begin{align}
    u_0= -60,\, u_1 = 0,\, u_2 = 12 \left(a_2+b_2+c_2\right), u_3 = 0, \, u_4 = -\left(a_2+b_2+c_2\right)^2,\, u_5 = 0
\end{align}
where $a_i$, $b_i$, $c_i$ are $i$-th symmetric polynomials of $\lambda_k$, $\mu_k$, $\nu_k$, respectively, 4d curve appears in the $\beta^6$ term. As before, an exact match with the curve \eqref{cc6} can be made by identifying $u_6$ with $z$ through a suitable shift.

\section{4d quantum curves}

The 4d quantum spectral curves of the rank-1 MN theories were derived in \cite{ACLccrgRankone} as follows. First, we know that the classical hamiltonian, $h=h(p,q)$, admits a natural quantization, $\hh=\hh(q, \hbar\frac{d}{dq})$, within the framework of complex crystallographic Calogero--Moser systems \cite{EFMV11ecm}. On the other hand, the classical spectral curves are given by the level sets $h^\vee(\alpha, k)=z$ of the ``dual'' classical hamiltonian. Hence, we can quantize these level sets by considering the eigenvalue problem 
\begin{equation}\label{qc}
    \hh^\vee(\alpha, \hbar\frac{d}{d\alpha})\psi=z\psi,\qquad \psi=\psi(\alpha;z)\,.
\end{equation}
Here $\hh^\vee$ is obtained from the quantum hamiltonian by replacing $q\mapsto \alpha$, $\frac{d}{dq}\mapsto \frac{d}{d\alpha}$, and replacing the coupling parameters by the dual parameters. 

The equation \eqref{qc} is an ODE on an elliptic curve $\E$, but due to its $\Z_m$-symmetry it can be converted to an ODE on the Riemann sphere $\mathbb P^1=\E/\Z_m$. To do this, one simply changes from $\alpha$ to the $\Z_m$-invariant coordinate $x=u(\alpha)$, as specified in Table \ref{table1}. The resulting ODE is a Fuchsian ODE of order $m$, with three singular points $e_0=\infty$, $e_1=0$, $e_2=1$. The corresponding ordinary differential operator can be written in a ``polynomial form'' that quantizes the polynomials $f=Q(x,y)-zP(x)$ used above to describe the elliptic pencils. The natural replacement for $y$, in view of \eqref{poisson}, is  
\begin{align}
  \hat{y}:= x(x-1)\hbar\frac{d}{dx}\,, \qquad m=3,4,6.
\end{align}
The quantum curve will be given, case by case, as a differential operator $F=F(x, \hat{y})$ with polynomial coefficients, following \cite{ACLccrgRankone}. In each case, the Fuchsian equation
\begin{equation}\label{qce}
 F(x, \hat{y})\phi=0\,,   
\end{equation}
has order $m$ and three singular points, $x=0,1,\infty$. The local monodromy data at $x=0$ and $x=1$ will be encoded in terms of parameters
$\mu_j$ and $\nu_j$, respectively, with $j=0,\dots, m-1$.\footnote{Compared to the notation in \cite{ACLccrgRankone}, our parameters $\mu_j$, $\nu_j$ are rescaled by a factor of $m$.} They are assumed normalized by 
\begin{equation}\label{zero}
    \sum_i\mu_i=\sum_i\nu_i=0\,.
\end{equation}
Additional parameters $\alpha_2,\dots, \alpha_m$ appearing in the formulas will be responsible for the local monodromy data at $x=\infty$.

\paragraph{Case $m=3$, type $E_6$:}
The quantum curve depends on parameters $\mu_{0,1,2}$ and $\nu_{0,1,2}$ subject to \eqref{zero}, and $\alpha_{2,3}$. It has the following form:
\begin{equation}\label{pqc3}
F= Y_2Y_1Y_0+{\alpha_2}x(x-1)Y_0+{(2\alpha_3x-z)}x(x-1)\,,
\end{equation}
where
\begin{equation}
    Y_j=\hat y-(\mu_j+\frac{j\hbar}{3})(x-1)-
    {(\nu_j+\frac{j\hbar}{3})}x\,.
\end{equation}
The equation \eqref{qce} in this case has order 3 and three regular singular points at $x=0, 1, \infty$. Note that a generic $3$rd order Fuchsian ODE with three singular points depends on $9$ parameters describing the local exponents at the singular points, and one accessory parameter. In our case, the local exponents at $x=0, 1$ are
$$
\hbar^{-1}\mu_j+\frac{j}{3} \quad(\text{at $x=0$})\,,\qquad \hbar^{-1}\nu_j+\frac{j}{3} \quad(\text{at $x=1$})\,,\qquad j=0,1,2\,.
$$
There are further three local exponents at $x=\infty$. 
However, due to the Fuchs relation, there are only two degrees of freedom for choosing them, corresponding to $\alpha_{2,3}$. The local exponents at $x=\infty$ are found by keeping the highest degree terms, $Y_j\sim x^2\hbar\frac{d}{dx}-(\mu_j+\nu_j)x$, and by acting on $x^{-\lambda}$. This gives the indicial equation
\begin{equation*}
(-\hbar(\lambda-2)+(\mu_2+\nu_2))(-\hbar(\lambda-1)+(\mu_1+\nu_1))(-\hbar\lambda +(\mu_0+\nu_0))+\alpha_2(-\hbar\lambda +(\mu_0+\nu_0))+2\alpha_3=0\,.     
\end{equation*}
We may present its roots (i.e. local exponents at $x=\infty$) in the form
$$
\hbar^{-1}\lambda_j+\frac{j}{3} \quad\text{with}\quad \lambda_0+\lambda_1+\lambda_2=0\,.
$$
Hence, \eqref{pqc3} is the general $3$rd order Fuchsian ODE with three singular points, normalised in such a way that the sum of local exponents equal $1$ at every singular point.  

We can define quantum mass parameters by
\begin{equation}
    \hat\lambda_j=\lambda_j+\frac{j\hbar}{3}\,,\quad \hat\mu_j=\mu_j+\frac{j\hbar}{3}\,,\quad \hat\nu_j=\nu_j+\frac{j\hbar}{3}\,.
\end{equation}
The quantum curve depends only on symmetric combinations of the quantum mass parameters, i.e.
\begin{equation}\label{qabc}
    a_i=\sigma_i(\hat\lambda_j), \quad b_i=\sigma_i(\hat\mu_j)\,,\quad c_i=\sigma_i(-\hat\nu_j)\,.
\end{equation}
Note that $a_1=b_1=-c_1= \hbar$, due to the normalisation \eqref{zero}. In the classical limit $\hbar\to 0$ these become the symmetric mass parameters $a_i, b_i, c_i$ used in the classical case.
The CB modulus, $z$, becomes the accessory parameter in the quantum curve. For generic parameters $\mu_j$, $\nu_j$ and $\alpha_{j}$, the local mondoromy around each singularity is semi-simple (i.e., diagonalisable). One says that it has spectral type $[1^3, 1^3, 1^3]$. 

\medskip

\paragraph{Case $m=4$, type $E_7$:} 
In this case the quantum curve depends on $11$ parameters $\mu_{0,1,2,3}$ and $\nu_{0,1,2,3}$ subject to \eqref{zero}, and $\alpha_{2,3,4}$. In addition, we specify that
\begin{equation}\label{rep4}
    \nu_0=\nu_2\,,\\ \nu_1=\nu_3\,.
\end{equation}
Hence, we have effectively $7$ mass parameters. The quantum curve has the form 
\begin{align}
F = Y_3Y_2Y_1Y_0+{\alpha_2}{x(x-1)}Y_1Y_0\label{pqc4}
+{2\alpha_3}{x(x-1)^2}Y_0+
    {(2\alpha_4(3x-1)-z)}{x(x-1)^2}\,,    
\end{align}
where
\begin{equation}
    Y_j=\hat y-(\mu_j+\frac{j\hbar}{4})(x-1)-(\nu_j+\frac{j\hbar}{2}){x}\,.
   \end{equation}
In that case, the Fuchsian equation \eqref{qce} has local exponents at $x=0,1$ given by
$$
\hbar^{-1}\mu_j+\frac{j}{4} \quad(\text{at $x=0$})\,,\qquad \hbar^{-1}\nu_j+\frac{j}{2} \quad(\text{at $x=1$})\,,\qquad j=0,1,2,3\,.
$$
Similarly to the $m=3$ case, the parameters $\alpha_{2,3,4}$ completely determine the local exponents at $x=\infty$ in the form 
$$
\hbar^{-1}\lambda_j+\frac{j}{4}\,,\quad\text{with}\quad \lambda_0+\dots+\lambda_3=0\,.
$$
Note that because of the condition \eqref{rep4}, we have two pairs of local exponents at $x=1$ which differ by an integer. In the theory of Fuchsian equations this is referred to as resonance. In general, in the presence of resonances one expects logarithmic terms in local solutions (and Jordan blocks in the local monodromy). However, in our case we insist on the absence of such terms, imposing that the local monodromy at $x=1$ remains semi-simple despite the resonances. Hence, the monodromy has spectral type $[1^4, 1^4, 2^2]$, with two pairs of repeated eigenvalues around $x=1$. As it turns out, this then fixes the ODE completely in the form \eqref{pqc4}, up to a single accessory parameter, $z$. The quantum mass parameters are
\begin{equation}
    \hat\lambda_j=\lambda_j+\frac{j\hbar}{4}\,,\quad \hat\mu_j=\mu_j+\frac{j\hbar}{4}\,,\quad \hat\nu_j=\nu_j+\frac{j\hbar}{2}\,.
\end{equation}
Here we only consider $\hat\nu_j$ for $j=0,1$, so the number of mass parameters is the same as in the classical case. The quantum curve depends on symmetric combinations $a_i, b_i, c_i$ of these quantum mass parameters, as defined in \eqref{qabc}.
   Note that in this case $a_1=b_1= 3\hbar/2$, $c_1= - \hbar/2$.

\medskip

\paragraph{Case $m=6$, type $E_8$:}
In this case the quantum curve depends on parameters $\mu_{0,\dots, 5}$ and $\nu_{0,\dots,5}$ subject to \eqref{zero}, and $\alpha_{2,3,4,5,6}$. In addition, we specify that
\begin{equation}\label{rep6}
    \mu_0=\mu_3\,,\ \mu_1=\mu_4\,\ \mu_2=\mu_5\,,\qquad \nu_0=\nu_2=\nu_4\,,\ \nu_1=\nu_3=\nu_5\,.
\end{equation}
Hence, we have effectively $8$ mass parameters. The quantum curve has the form 
\begin{align}
       F= & Y_5 Y_4 Y_3 Y_2 Y_1 Y_0
     +   {\alpha_2}{x\left(x-1\right)} Y_3 Y_2 Y_1 Y_0\notag
     +   {2\alpha_3}{x\left(x-1\right)^2} Y_2 Y_1 Y_0 \notag\\ 
     + &  {6\alpha_4}{x^2 \left(x-1\right)^2} Y_1 Y_0\label{pqc6}
     + {24\alpha_5}{x^2 \left(x-1\right)^3} Y_0 
     +   (24\alpha_6(5 x-2)-z){x^2 \left(x-1\right)^3}\,,
\end{align}
where
\begin{align}
    Y_j = \hat y-({\mu}_j+\frac{j\hbar}{3}){(x-1)} -({\gamma}_j+\frac{j\hbar}{2}){x}\,.
\end{align}
In that case, the Fuchsian equation \eqref{qce} has local exponents at $x=0,1$ given by
\begin{align*}
\hbar^{-1}\mu_j+\frac{j}{3} \quad(\text{at $x=0$})\,,\qquad \hbar^{-1}\nu_j+\frac{j}{2} \quad(\text{at $x=1$})\,,\qquad j=0,\dots, 5\,.
\end{align*}
The parameters $\alpha_{2,\dots, 6}$ completely determine the local exponents at $x=\infty$ in the form 
\begin{align*}
  \hbar^{-1}\lambda_j+\frac{j}{6}\,,\quad\text{with}\quad \lambda_0+\dots+\lambda_5=0\,.  
\end{align*}

Note that because of the conditions \eqref{rep6}, at $x=0$ we have three pairs of local exponents whose difference is an integer, and at $x=1$ we have two triples of local exponents whose pairwise differences are integers. Hence, this is a highly resonant case. Again, we insist on the local monodromy at $x=0, 1$ being semi-simple despite the resonances. Hence, the monodromy has spectral type $[1^6, 2^3, 3^2]$ in such case. As it turns out, this then fixes the ODE completely in the form 
\eqref{pqc6}, with a single accessory parameter, $z$.    
The quantum mass parameters are
\begin{equation}
    \hat\lambda_j=\lambda_j+\frac{j\hbar}{6}\,,\quad \hat\mu_j=\mu_j+\frac{j\hbar}{3}\,,\quad \hat\nu_j=\nu_j+\frac{j\hbar}{2}\,.
\end{equation}
Here we only take $\hat\mu_{0,1,2}$  and $\hat\nu_{0,1}$, so the number of mass parameters is the same as in the classical case. The quantum curve depends only on symmetric combinations $a_i, b_i, c_i$ of the quantum mass parameters, as defined in \eqref{qabc}. Note that in this case $a_1= 5\hbar/2$, $b_1= \hbar$, $c_1= - \hbar/2$.

\section{5d quantum curves and 4d limits}
The quantum curves for Seiberg's $E_{6,7,8}$ theories have been studied in \cite{Moriyama2020_SpectralTSDelPezzo, MoriyamaY2021_AffineWeylQuantumCurve}. In particular, the derivation in \cite{MoriyamaY2021_AffineWeylQuantumCurve} was based on a rectangular realization and a $q$-variant of the Frobenius method. Below we use a similar approach for a triangular realization \cite{Moriyama2020_SpectralTSDelPezzo}. This gives a more systematic method compared to the original derivation in \cite{Moriyama2020_SpectralTSDelPezzo}, and makes a clear parallel with our characterization of the 4d quantum curves. 

\subsection{5d quantum curves}

A natural quantization of $\C^*\times \C^*$ is the quantum torus,
\begin{align}
    \mathcal{A} = \C\langle \hat{t}^{\pm1}, \hat{w}^{\pm1}\rangle/\{\hat{t} \hat{w} = q \hat{w} \hat{t} \}, 
\end{align}
so each quantum 5d curve will be represented by an element $G$ of $\mathcal{A}$ of the form
\begin{align}
    G = \sum_{\substack{p \ge 0, q \ge 0 \\ p+ q \le m }} c_{p,q} \hat{t}^p \hat{w}^q\,,\qquad m=3,4,6.
\end{align}
The algebra $\mathcal{A}$ can be realized (in many ways) as an algebra of $q$-difference operators acting on Laurent polynomials. The standard way is to take a left ideal $\mathcal{A}(\hat w-1)\subset \mathcal{A}$ and consider the left $\mathcal{A}$-module $\mathcal{M}:=\mathcal{A}/\mathcal{A}(\hat w-1)$. The mapping $\hat t^a\hat w^b\mapsto t^a$ identifies $\mathcal{M}$ with $\C[t, t^{-1}]$, with the (left) $\mathcal{A}$-action given by $(\hat{t} f)(t)  =  t f(t)$ and $(\hat{w} f)(t) = f(q^{-1} t)$ for $f\in\C[t, t^{-1}]$. In addition to this, we will need two further realizations. Namely, consider three left $\mathcal{A}$-modules with induced $\mathcal{A}$-actions as follows:
\begin{align}
    \mathcal{M}_1 & = \mathcal{A}/\mathcal{A}(\hat{w} -1) \sim \mathbb{C}[t, t^{-1}], & (\hat{t} f)(t) & =  t f(t), (\hat{w} f)(t) = f(q^{-1} t)\\
    \mathcal{M}_2 & = \mathcal{A}/\mathcal{A}(\hat{t} -1) \sim \mathbb{C}[w, w^{-1}], & (\hat{t} f)(w) & = f(q t), (\hat{w} f)(w) = w f(w)\\
    \mathcal{M}_3 & = \mathcal{A}/\mathcal{A}(\hat{w}\hat{t}^{-1}-1)\sim \mathbb{C}[t, t^{-1}], &  (\hat{t} f)(t) & = t f(t), (\hat{w} f)(t) = q^{-1}t f(q^{-1}t). 
\end{align}
The first of these has been explained already; $\mathcal{M}_2$ is constructed similarly, reversing the roles of $\hat t$ and $\hat w$. For $\mathcal{M}_3$, the identification 
$\mathcal{M}_3\sim \C[t, t^{-1}]$ may be chosen\footnote{The choice of the identification is not unique, but the resulting $\mathcal{A}$-modules are isomorphic.} as $\hat t^a\hat w^b\mapsto q^{-b(b+1)/2}t^{a+b}$.
Our choice of these three $A$-modules reflects the triangular structure of the five-brane web used for the construction of the classical 5d curves. 

Now, for each of the three realizations, we are going to consider the $q$-difference equation
\begin{equation}\label{qce5}
    G\psi=0
\end{equation}
and analyse its solutions by a formal series, near $0$ or $\infty$. This can be viewed as a $q$-variant of the Frobenius method for ODEs, cf. \cite{MoriyamaY2021_AffineWeylQuantumCurve, takemura2018q}.

\begin{enumerate}
    \item For $\mathcal{M}_1$, we consider a formal power series in $t$ of the form
\begin{align}
    \psi(t) = \sum_{j \ge 0} c_j t^{\rho + j }, \qquad \rho\in\C\,,
\end{align}
and rewrite $G$ as
\begin{align}
    G = \sum_{i = 0}^m \hat{t}^i a_i(\hat{w})\,.
\end{align}
Then \eqref{qce5} gives
\begin{align}
    0 = G\psi(t) =\sum_{k \ge 0} \left(\sum_{j=0}^k c _j a_{k-j} ( q^{-\rho - j } ) \right) t^{\rho + k}\,. 
\end{align}
This gives a system of recurrence relations for $\{c_j\}$. 

\item For $\mathcal{M}_2$, we consider a formal power series in $w$ of the form
\begin{align}
    \psi(w) = \sum_{j \ge 0} c_j w^{\rho + j }
\end{align}
and rearrange $G$ as
\begin{align}
    G = \sum_{i = 0}^m \hat{w}^i b_i(\hat{t})\,.
\end{align}
Then \eqref{qce5} gives
\begin{align}
    0 = G\psi(w)= \sum_{k \ge 0} \left(\sum_{j=0}^k c _j b_{k-j} ( q^{\rho + j } ) \right) w^{\rho + k} 
\end{align}
and a system of recurrence relations for $\{c_j\}$.

\item Finally, for $\mathcal{M}_3$ we consider a formal power series in $t^{-1}$\, \footnote{The reason for considering $t^{-1}$ is that in the classical case we need to work near $t,w\sim \infty$.} 
\begin{align}
    \psi(t) = \sum_{j \ge 0} c_j t^{-\rho - j }\,.
\end{align}
It is more convenient to work with $\widetilde G:=t^{-m}G$ and rewrite it in terms of $\hat t^{-1}$ and $\hat{v} := \hat{w} \hat{t}^{-1}$, 
\begin{align}
    \widetilde G=\sum_{\substack{a \ge 0, b \ge 0 \\ a+ b \le m }} c_{a,b} \hat{t}^{a-m}\hat{w}^b=\sum_{\substack{a \ge 0, b \ge 0 \\ a+ b \le m }} q^{-b(b+1)/2}c_{a,b} \hat{t}^{a+b-m}\hat{v}^b=
    \sum_{i = 0}^{m} \hat{t}^{-i} d_i(\hat{v})\,, \qquad {\rm deg}\, d_i \le i.
\end{align}
Then \eqref{qce5} gives
\begin{align}
    0 =\widetilde G \psi(t)  =\sum_{k \ge 0} \left(\sum_{j=0}^k {c}_j d_{k-j} ( q^{\rho + j } ) \right) t^{-\rho - k}\,.
\end{align}
\end{enumerate}

Analogous to the differential case, the roots of $a_0(w) =0 $ for $\mathcal{M}_1$, $b_0(t) = 0$ for $\mathcal{M}_2$ and $d_0(v) =0 $ for $\mathcal{M}_3$ are the local exponents. The formal local series solutions could be rendered invalid whenever there are exponents differing by $q^n$ with some integer $n$. This is referred to as resonance. In general, it could force logarithmic terms, as in the case of ODEs. However, sometimes the logarithmic terms do not appear despite a resonance. 

Below we are going to impose certain conditions on $G$, allowing some resonances but insisting on the absence of logarithms in the formal solutions. In doing so we will rely on the following result which gives conditions for the absence of logarithms in presence of resonance.
\begin{prop}[Proposition 3.1, {\cite{MoriyamaY2021_AffineWeylQuantumCurve}}]\label{prop:5.1}
For a difference operator $D = \sum_{i=0}^d x^iA_i(y)$, we have

$(1)$ $D$ has non-logarithmic singularities at $x = 0$ with $y = a,qa,\dots,
q^{m-1}a$ iff $A_i(y)\propto \prod_{j=0}^{m-i-1}(y-q^ja)$ for $0\le i\le m-1$,

$(2)$ $D$ has non-logarithmic singularities at $x = \infty$  with $y = a,q^{-1}a,\dots,q^{-m+1}a$ iff $A_i(y)\propto 
\prod_{i=0}^{m-i-1}(y-q^{-j}a)$ for $d-m+1\le i\le d$.
\end{prop}
As we will see, the conditions of the non-logarithmic resonance completely fix $G$, leaving one parameter undetermined. This accessory parameter is then identified as the CB modulus $U$.  In the following, we will use the notation $\hat{L}_i, \hat{M}_i, \hat{N}_i$ for the mass parameters, keeping the same notation $\chi_j^\lambda$, $\chi_j^\mu$, $\chi_j^\nu$ for their $j$-th elementary symmetric polynomials, that we used in the classical case.

\paragraph{Case $m=3$, type $E_6$:}
We impose the following conditions: 
\begin{itemize}
    \item[$\ast$] For $\mathcal{M}_1$, it is
    \begin{align}
        a_0(w) = c_{0,3} \prod_{i=0}^{2} (w - \hat{M}_i)
    \end{align}
    \item[$\ast$] For $\mathcal{M}_2$, it is
    \begin{align}
        b_0(t) = c_{3,0} \prod_{i=0}^{2} (t- \hat{N}_i)
    \end{align}
        \item[$\ast$] For $\mathcal{M}_3$, it is
    \begin{align}
        d_0(v) = \frac{1}{q^3} c_{0,3} \prod_{i=0}^{2} (v+ \hat{L}_i)
    \end{align}
\end{itemize}
This fixes the exponents for $\mathcal{M}_1, \mathcal{M}_2, \mathcal{M}_3$ to be $\hat{M}_i$, $\hat{N}_i$, and $\hat{L}_i$, respectively. For consistency, we have to impose the condition
\begin{align}
    q^3 \chi_3^{\mu} = \chi_3^\lambda \chi_3^\nu  
\end{align}
on the mass parameters. 
By rescaling, we may assume
\begin{align}
    c_{0,3} = q, \;\; c_{3,0} = q^{-1}, \;\; c_{0,0} =  -1, \;\; \chi_3^\lambda = q, \;\; \chi_3^\mu = q^{-1}, \;\; \chi_3^{\nu} = q 
\end{align}
Finally, we can solve for the other coefficients of the quantum curve $G$:
\begin{align}
    c_{0,1}=q \chi _2^{\mu },\, c_{0,2}=-q \chi _1^{\mu }, \, c_{1,0}=\frac{\chi _2^{\nu }}{q},\, c_{1,2}=\frac{\chi _1^{\lambda }}{q}, \, c_{2,0}=-\frac{\chi _1^{\nu }}{q}, \,c_{2,1}=\frac{\chi _2^{\lambda }}{q^2}
\end{align}
The coefficient $c_{1,1}$ remains undetermined and will be identified with the CB modulus $U$. 

\paragraph{Case $m=4$, type $E_7$:}
We impose the conditions as follows:
\begin{itemize}
    \item[$\ast$] For $\mathcal{M}_1$, it is
    \begin{align}
        a_0(w) = c_{0,4} \prod_{i=0}^{3} (w - \hat{M}_i)
    \end{align}
    \item[$\ast$] For $\mathcal{M}_2$, it is
    \begin{align}
        b_0(t) & = c_{4,0} \prod_{i=0,1} (t- \hat{N}_i)(t- q \hat{N}_i) \notag \\
        b_1(t) & \propto \prod_{i=0,1} (t- \hat{N}_i)
    \end{align}
        \item[$\ast$] For $\mathcal{M}_3$, it is
    \begin{align}
        d_0(v) = \frac{1}{q^6} c_{0,4} \prod_{i=0}^{3} (v+ \hat{L}_i)
    \end{align}
\end{itemize}
This fixes exponents for $\mathcal{M}_1, \mathcal{M}_2, \mathcal{M}_3$. Note that we have two resonant
pairs of exponents for $\mathcal{M}_2$, in parallel to the 4d $m=4$ case. The conditions imposed on $b_1$ guarantee the absence of logarithms, cf. Prop. \ref{prop:5.1}.
For consistency, we have to impose the condition
\begin{align}
    q^4 \chi_4^{\mu} = \chi_4^\lambda (\chi_2^\nu )^2
\end{align}
By rescaling, we may assume
\begin{align}
c_{0,4}=q^{\frac{3}{2}},\;\; c_{4,0}=q^{-3},\;\; c_{0,0}=1, \;\; \chi_4^{\lambda }=q^{\frac{3}{2}}, \;\; \chi_4^{\mu }=q^{-\frac{3}{2}},\;\; \chi_2^{\nu }=q^{\frac{1}{2}}
\end{align}
Finally, we can solve for the other coefficients:
\begin{align}
    \begin{split}
       & c_{0,1}=-q^{3/2} \chi _3^{\mu },\, c_{0,2}=q^{3/2} \chi _2^{\mu }, \, c_{0,3}=-q^{3/2} \chi _1^{\mu }, \, c_{1,0}=-\frac{(q+1) \chi_1^{\nu }}{q^{3/2}},\\
       & c_{1,1}=\frac{\chi _3^{\lambda }}{q^2}+\chi _3^{\mu } \chi _1^{\nu },\,c_{1,3}=\frac{\chi _1^{\lambda }}{q^{3/2}},\,c_{2,0}=\frac{q^2+\sqrt{q} \left(\chi _1^{\nu }\right){}^2+1}{q^{5/2}}, \\ 
       & c_{2,1}=-\frac{q^{5/2} \chi _3^{\mu }+\chi _3^{\lambda } \chi _1^{\nu }}{q^{7/2}},\,c_{2,2}=\frac{\chi _2^{\lambda }}{q^{7/2}},\,c_{3,0}=-\frac{(q+1) \chi _1^{\nu }}{q^3},\,c_{3,1}=\frac{\chi _3^{\lambda }}{q^{9/2}}
\end{split}
\end{align}
In this case, the coefficient $c_{1,2}$ remains undetermined and is identified with CB modulus $U$. 

\paragraph{Case $m=6$, type $E_8$:}
We impose the conditions as follows:
\begin{itemize}
    \item[$\ast$] For $\mathcal{M}_1$, it is
    \begin{align}
    \begin{split}
        a_0(w) & = c_{0,6} \prod_{i=0}^{2} (w - \hat{M}_i) (w - q^{-1} \hat{M}_i) \\
        a_1(w) & \propto \prod_{i=0}^{2} (w - \hat{M}_i) 
    \end{split}
    \end{align}
    \item[$\ast$] For $\mathcal{M}_2$, it is
    \begin{align}
    \begin{split}
        b_0(t) & = c_{6,0} \prod_{i=0,1} (t- \hat{N}_i)(t- q \hat{N}_i) (t- q^2 \hat{N}_i)  \\
        b_1(t) & \propto \prod_{i=0,1} (t- \hat{N}_i) (t- q \hat{N}_i)  \\ 
        b_2(t) & \propto \prod_{i=0,1} (t- \hat{N}_i)
    \end{split}
    \end{align}
        \item[$\ast$] For $\mathcal{M}_3$, it is
    \begin{align}
        c_0(v) = \frac{1}{q^{15}} c_{0,6} \prod_{i=0}^{5} (v+ \hat{L}_i)
    \end{align}
\end{itemize}
Again, these conditions fix the exponents for each of $\mathcal{M}_1, \mathcal{M}_2, \mathcal{M}_3$, and ensure that there are no logarithmic terms, despite resonances. For consistency, we have to impose the condition
\begin{align}
    q^6 (\chi_3^{\mu})^3= \chi_6^\lambda (\chi_2^\nu )^3
\end{align}
By rescaling, we may assume
\begin{align}
c_{0,6}=q^{5},\;\; c_{6,0}=q^{-\frac{15}{2}},\;\; c_{0,0}=1, \;\; \chi_6^{\lambda }=q^{\frac{5}{2}}, \;\; \chi_3^{\mu }=q^{-1},\;\; \chi_2^{\nu }=q^{\frac{1}{2}}
\end{align}
Finally, we can solve for the other coefficients:
\begin{align}
\begin{split}
    & c_{0,1}=-q (q+1) \chi _2^{\mu },\, c_{0,2}=\left(\chi _2^{\mu }\right){}^2 q^3+\left(q^3+q\right) \chi _1^{\mu },\, c_{0,3}=-q (q+1) \left(\chi _1^{\mu } \chi _2^{\mu } q^2+q^2-q+1\right),\\
    & c_{0,4}=q^3 \left(q \left(\chi _1^{\mu }\right){}^2+\left(q^2+1\right) \chi _2^{\mu }\right),\, c_{0,5}=-q^4 (q+1) \chi _1^{\mu },\,c_{1,0}=-\frac{\left(q^2+q+1\right) \chi _1^{\nu }}{q^{5/2}},\\
    & c_{1,1}=\frac{(q+1)^2 \chi _1^{\nu } \chi _2^{\mu } q^{3/2}+\chi _5^{\lambda }}{q^3},\,c_{1,2}=-\frac{q \chi _1^{\lambda }+\left(q^2+q+1\right) \sqrt{q} \chi _1^{\mu } \chi _1^{\nu }+\chi _2^{\mu } \left(\chi _1^{\nu } \chi _2^{\mu } q^{5/2}+\chi _5^{\lambda }\right)}{q^2},\\
    & c_{1,3}=\chi _1^{\lambda } \chi _2^{\mu }+\frac{\chi _1^{\nu } \left(\chi _1^{\mu } \chi _2^{\mu } q^2+q^2+q+1\right)}{q^{3/2}}+\frac{\chi _1^{\mu } \chi _5^{\lambda }}{q^2},\,c_{1,4}=-\frac{\chi _1^{\nu } \chi _2^{\mu } q^{5/2}+\chi _1^{\lambda } \chi _1^{\mu } q^2+\chi _5^{\lambda }}{q^2},\\
    & c_{1,5}=\chi _1^{\lambda },\,c_{2,0}=\frac{\left(q^2+q+1\right) \left(q^2-q+\sqrt{q} \left(\chi _1^{\nu }\right){}^2+1\right)}{q^{9/2}},\\
    & c_{2,1}=-\frac{(q+1) \left(\left(\left(\chi _1^{\nu }\right){}^2 q^{5/2}+q^4+q^2\right) \chi _2^{\mu }+\chi _1^{\nu } \chi _5^{\lambda }\right)}{q^{11/2}},\\
    & c_{2,2}=\frac{\left(\chi _2^{\mu }\right){}^2 q^3+\chi _1^{\lambda } \chi _1^{\nu } q+\chi _1^{\mu } \left(\left(\chi _1^{\nu }\right){}^2 q^{3/2}+q^3+q\right)+\chi _4^{\lambda }+\chi _1^{\nu } \chi _2^{\mu } \chi _5^{\lambda }}{q^{9/2}},\\
    & c_{2,4}=\frac{\chi _2^{\lambda }}{q^4},\,c_{3,0}=-\frac{\chi _1^{\nu } \left(\left(\chi _1^{\nu }\right){}^2 q^{3/2}+q^4+2 q^3+2 q+1\right)}{q^6}, \\
    & c_{3,1}=\frac{\chi _1^{\nu } \chi _2^{\mu } (q+1)^2}{q^5}+\frac{\left(\chi _1^{\nu }\right){}^2 \chi _5^{\lambda }}{q^7}+\frac{\left(q^2+1\right) \chi _5^{\lambda }}{q^{15/2}},\\
    & c_{3,2}=-\frac{\chi _1^{\lambda } q^{3/2}+\chi _1^{\mu } \chi _1^{\nu } q^2+\chi _1^{\nu } \chi _4^{\lambda }+\sqrt{q} \chi _2^{\mu } \chi _5^{\lambda }}{q^7},\,c_{3,3}=\frac{\chi _3^{\lambda }}{q^7}, \\
    & c_{4,0}=\frac{\left(q^2+q+1\right) \left(q^2-q+\sqrt{q} \left(\chi _1^{\nu }\right){}^2+1\right)}{q^7},\,c_{4,1}=-\frac{(q+1) \left(\chi _2^{\mu } q^3+\chi _1^{\nu } \chi _5^{\lambda }\right)}{q^9},\\
    & c_{4,2}=\frac{\chi _4^{\lambda }}{q^9},c_{5,0}=-\frac{\left(q^2+q+1\right) \chi _1^{\nu }}{q^{15/2}},\,c_{5,1}=\frac{\chi _5^{\lambda }}{q^{10}}
\end{split}
\end{align}
The coefficient $ c_{2,3} $ remains undetermined and is identified with the CB modulus $U$.

\begin{remark}
In the classical limit $\hbar \to 0$ (and thus $q \to 1$), it is straightforward to verify that the 5d quantum curve $G$ turns into the classical 5d curve $g$.  
\end{remark}

\begin{remark}
Our expressions for the quantum curves $G$ seem to agree with the results in \cite{Moriyama2020_SpectralTSDelPezzo}, although we have not done a full comparison due to differences in the setup and notation.   
\end{remark}

\subsection{4d limit}

Inspired by the classical case \eqref{clift_1}, we uplift $t$ and $w$ to quantum variables
\begin{align}\label{qlift_1}
    \hat{t } = x e^{\beta \hbar x (x-1) \frac{d}{dx}}, \quad  \hat{w } = (1- x) e^{\beta \hbar x (x-1) \frac{d}{dx}}
\end{align}
One can show that $\hat{t} \hat{w} = q \hat{w} \hat{t}$ with $q=e^{\beta \hbar}$. 
To see this, one uses that $T:=e^{\gamma(x^2 - x )\frac{d}{dx}}$ acts on a function $f(x)$ by
\begin{align}
    Tf(x) = e^{\gamma x(x - 1 ) \frac{d}{dx}} f(x) = f\left(\frac{x}{x + e^{\gamma} (1-x)}\right).
\end{align}
This is seen by ``straightening'' the vector field $\gamma x(x - 1 ) \frac{d}{dx}$ by a change of variables,
\begin{equation}
    \gamma x(x-1)\frac{d}{dx} =\frac{d}{du}\qquad\text{for}\quad u=\gamma^{-1}\log(1-\frac{1}{x})\,,\quad x=\frac{1}{1-e^{\gamma u}}\,.
\end{equation}
Hence,  
\begin{align}
    \frac{\hat{t} \hat{w} f(x)}{\hat{w} \hat{t} f(x)} = \frac{x T (1-x) T f(x)}{(1-x) T x T f(x)} = \frac{x (1- \frac{x}{x + e^{\gamma} (1-x)}) T^2 f(x)}{(1-x) \frac{x}{x + e^{\gamma} (1-x)} T^2 f(x)} = e^\gamma. 
\end{align}
We now substitute \eqref{qlift_1} into the 5d quantum curve $G$, and expand in series in $\beta$, using Mathematica. In doing so, we also replace the CB modulus, $U$, by \eqref{Uexp}. By tuning the parameters $u_i$, we can achieve that the expansion of $G$ will have the first nonzero term at the $m$-th order in $\beta$. This term is then compared with the appropriate 4d quantum curve.
In each case below, we choose the mass parameters in the form  
\begin{align}
    \hat{L}_i = e^{\beta \hat{\lambda}_i}, \quad \hat{M}_i = e^{-\beta \hat{\mu}_i}, \quad \hat{N}_i = e^{\beta \hat{\nu}_i}\,,
\end{align}
and denote by $a_i, b_i, c_i$ the elementary symmetric polynomials of $\hat \lambda_i$, $\hat \mu_i$ and $\hat \nu_i$, respectively.

\paragraph{Case $m=3$, type $E_6$:}
In this case, we choose
\begin{align}
    u_0\to -6,\quad u_1\to 3 \hbar, \quad u_2\to a_2+b_2+c_2-\frac{3 \hbar ^2}{2}
\end{align}
With this choice, the $\beta^3$-term in the expansion of $G$ matches the 4d $m=3$ quantum curve $F=F(x, \hat y)$.

\paragraph{Case $m=4$, type $E_7$:}
In this case, we choose
\begin{align}
    u_0 = -12,\quad u_1 = 12 \hbar, \quad u_2 = 2 \left(a_2+b_2+c_2\right)-9 \hbar ^2, \quad u_3 = -2 \hbar  \left(a_2+b_2+c_2\right) + 5 \hbar^3. 
\end{align}
With this choice, the $\beta^4$-term in the expansion of $G$ matches the 4d $m=4$ quantum curve $F=F(x, \hat y)$.

\paragraph{Case $m=6$, type $E_8$:}
In this case, we choose
\begin{align}
\begin{split}
  &  u_0 =  -60, \, u_1 = 180 \hbar , \\
  &  u_2\to 12 \left(a_2+b_2+c_2\right)-\frac{607 \hbar ^2}{2}, \, u_3 = -36 \hbar  \left(a_2+b_2+c_2\right) +  \frac{741 \hbar ^3}{2}, \\
  & u_4 = -\left(a_2+b_2+c_2\right)^2 + \frac{119}{2} \hbar ^2 \left(a_2+b_2+c_2\right)-\frac{34639 \hbar ^4}{96}, \\
  & u_5 = 3 \hbar  \left(a_2+b_2+c_2\right){}^2 -\frac{141}{2} \hbar ^3 \left(a_2+b_2+c_2\right)+\frac{9439 \hbar ^5}{32}.
\end{split}
\end{align}
With this choice, the $\beta^6$-term in the expansion of $G$ matches the 4d $m=6$ quantum curve $F=F(x, \hat y)$.

\begin{remark}
 Note that the quantum uplift of \eqref{clift_2} is
\begin{align} \label{qlift_2}
    \hat{t} = x, \quad \hat{w}= (1-x) e^{\beta \hbar x \frac{d}{dx}},
\end{align}
which also satisfies $\hat{t}\hat{w}=q \hat{w}\hat{t}$. 
This produces the same results as \eqref{qlift_1} in the 4d limit.    
\end{remark}

\section{Summary and outlook}

In this paper, we performed a comparative study of classical and quantum curves of some 4d and 5d SCFTs, namely, the rank-1 Minahan--Nemeschansky and Seiberg's theories of type $E_{6,7,8}$. For the 4d theories, we used the recently found presentations \cite{ACLccrgRankone}, different from the previously known Weierstrass models. To match them with the 5d curves \cite{KimYagi14, Moriyama2020_SpectralTSDelPezzo}, we use five-brane web diagrams of a triangular type. This makes the properties of 4d and 5d curves very similar, and allows us to make a direct comparison by taking a 4d limit. In particular, inspired by the classical cases \eqref{clift_1} and \eqref{clift_2}, we examined two different 4d limits \eqref{qlift_1} and \eqref{qlift_2}, both of which yield the same results for the quantum curves. This raises questions about the potential existence of additional 4d limits and more systematic methods for identifying them. Specifically, the 4d limit \eqref{qlift_2} can be understood in terms of the Hanany-Witten transition from the triangular web to the rectangular web\footnote{We thank S-S.~Kim and F.~Yagi for their insightful discussion on this topic.}.

It would be interesting to apply a similar approach to the higher rank versions of these theories. For the higher rank Minahan--Nemeschansky theories, a nice way of doing this is based on the link to the elliptic Calogero--Moser systems of complex-crystallographic type \cite{ACL2}. These results suggest natural candidates for the classical and quantum curves of the higher rank 5d theories. 

Another interesting question is about the relation of the 5d theories to integrable systems. Based on the rank-one studies in \cite{BershteinGM_ClusterQPainleve, Mizuno_qPClusterToric}, one may expect a relation to suitable cluster integrable systems, as well as to $q$-Painlev\'e systems, in a higher rank. 

We also expect further examples of elliptic integrable systems that can be identified with SW integrable systems of 4d SCFTs. This may allow us to make predictions for the spectral curves of the relevant 5d theories.

\section*{Acknowledgement}
We would like to thank Philip Argyres, Sung-Soo Kim, Kimyeong Lee, Sanefumi Moriyama, Futoshi Yagi, and Yasuhiko Yamada for valuable discussions. 
YL is supported by KIAS individual grant PG084801.

\bibliographystyle{myJHEP}
\bibliography{mybibs5d}

\end{document}